\documentclass[acmsmall, screen]{acmart}
\usepackage{subcaption}
\usepackage{pbox}
\newcommand{\jie}[1]{{\color{red} {#1}}}
\AtBeginDocument{%
  \providecommand\BibTeX{{%
    \normalfont B\kern-0.5em{\scshape i\kern-0.25em b}\kern-0.8em\TeX}}}

\setcopyright{acmcopyright}
\copyrightyear{2018}
\acmYear{2018}
\acmDOI{10.1145/1122445.1122456}

\acmConference[Woodstock '18]{Woodstock '18: ACM Symposium on Neural
  Gaze Detection}{June 03--05, 2018}{Woodstock, NY}
\acmBooktitle{Woodstock '18: ACM Symposium on Neural Gaze Detection,
  June 03--05, 2018, Woodstock, NY}
\acmPrice{15.00}
\acmISBN{978-1-4503-XXXX-X/18/06}

\begin{document}

\title[Stress Reduction for Online Content Moderators]{Awe Versus Aww: The Effectiveness of Two Kinds of Positive Emotional
Stimulation on Stress Reduction for Online Content Moderators}


\author{Christine L. Cook}
\affiliation{%
  \institution{National Chengchi University}
  \city{Taipei City}
  \country{Taiwan}}
\email{christinelcook@outlook.com}

\author{Jie Cai}
\affiliation{%
  \institution{New Jersey Institute of Technology}
  \city{Newark}
  \country{USA}}
\email{jie.cai@njit.edu}

\author{Donghee Yvette Wohn}
\affiliation{%
  \institution{New Jersey Institute of Technology}
  \city{Newark}
  \country{USA}}
\email{yvettewohn@gmail.com}

\renewcommand{\shortauthors}{Cook et al.}


\begin{abstract}
When people have the freedom to create and post content on the internet, particularly anonymously, they do not always respect the rules and regulations of the websites on which they post, leaving other unsuspecting users vulnerable to sexism, racism, threats, and other unacceptable content in their daily cyberspace diet. However, content moderators witness the worst of humanity on a daily basis in place of the average netizen. This takes its toll on moderators, causing stress, fatigue, and emotional distress akin to the symptomology of post-traumatic stress disorder (PTSD). The goal of the present study was to explore whether adding positive stimuli to breaktimes–-images of baby animals or beautiful, aweinspiring landscapes–-could help reduce the negative side-effects of being a content moderator. To test this, we had over 300 experienced content moderators read and decide whether 200 fake text-based social media posts were acceptable or not for public consumption. Although we set out to test positive emotional stimulation, however, we actually found that it is the cumulative nature of the negative emotions that likely negates most of the effects of the intervention: the longer the person had practiced content moderation, the stronger their negative experience. Connections to compassion fatigue and how best to spend work breaks as a content moderator are discussed.
\end{abstract}


\maketitle
\jie{Preprint accepted by CSCW 2022.}
\section{Introduction}



In the toxic world of cyberspace, online content moderators are the protectors of netizens everywhere. As either an employee or a volunteer, content moderators sift through everything people have posted or have reported as being inappropriate for hours on end \cite{Seering2019ModeratorAlgorithms}, seeing the worst content people on the internet have to offer: sexism, racism, and more, with varying degrees of intensity \cite{Wohn2019VolunteerExperience}. Much like police officers working on internet crime teams (see \cite{Krause2009IdentifyingInvestigators}), consistent exposure to this kind of content can lead to severe levels of stress and fatigue among content moderators \cite{Dang2018ButModeration}. Although this is problematic in and of itself, the most unsettling part of this issue is the fact that the problem is only poised to grow over time. Content moderation is a growing field that upholds the social media infrastructure of today, meaning that more and more people will be employed over time to do this job and be subjected to this kind of suffering (see \cite{Carmi2019TheModerators, Parks2019DirtyCleaners}). As we continue to produce more content, more people will have to filter through it and endure the ill-effects thereof. Content moderators are quickly becoming a major vulnerable group in cyberspace.

However, researchers have begun to take interest in the wellbeing of content moderators \cite{Steiger2021TheModerators,Roberts2016CommercialWork, Newton2019TheAmerica} and have examined the effectiveness of current methods to ease the burden on content moderators (e.g., \cite{Cai2019WhatPerspectives,Seering2019DesigningBehaviors,Seering2017ShapingExample-setting,Chandrasekharan2019CrossMod:Moderators}), referring to both moderators in volunteer and commercial content moderation, as they both experience exposure to harmful content and suffer from emotional labors \cite{Steiger2021TheModerators}. Researchers have tried to reduce the burden on content moderators by altering the offensive content to reduce exposure, often by blurring image and video content \cite{Dang2018ButModeration, Das2020FastContent}, and by either supporting or replacing human moderators with artificial intelligence systems, to varying degrees of success \cite{Ruckenstein2020Re-humanizingCare}. However, psychological studies would suggest that we still need interventions that target the moderator themselves, helping them to cope with the emotions and distress that accompanies their work (e.g., \cite{Benjelloun2020PsychologicalReport}). The present study takes inspiration from the very internet community the moderators aim to protect; one popular way to reduce stress and increase mental wellbeing on the internet is to actively seek out positive content to consume (e.g., \cite{Bagliere2020ScienceCNN,Dockrill2016JustLevels}). Two of the most purportedly effective sources of this positivity are baby animal images \cite{Bagliere2020ScienceCNN} and beautiful landscapes \cite{Dockrill2016JustLevels}. Thus, in this study, we aim to test the two associated positive emotions–-cuteness and awe–-for their therapeutic effects on stress and fatigue in the content moderating context to see if they can work for more acutely distressed emotional laborers the way they appear to for the average netizen.

Both cuteness and awe have been shown to be connected to increased levels of prosociality (e.g., \cite{Guan2019AweTendency,Sherman2013IndividualCareful} ) and a sense of connectedness to others (e.g., \cite{Bai2017AweSelf, Sherman2011CutenessEmotion}), but they also have their own distinct benefits that could impact content moderators differently. For instance, cuteness has been repeatedly connected to carefulness, attentional control, and caretaking behaviours \cite{Glocker2009BabyAdults, Karreman2020ExposureMothers, Nittono2012TheFocus}. This could impact content moderation in several different ways. The extra carefulness could be linked to a reduction in speed, but an increase in accuracy, while increased attention control has the potential to increase productivity by reducing distraction. From an emotional perspective, cuteness has been linked to stress reduction \cite{TourismWesternAustralia2020Study:Singapore} and, when paired over time with another stimulus, has also been shown to increase the quality of positive experience with paired stimuli \cite{McNulty2017AutomaticConditioning}. For content moderators, adding cute stimuli to break up the negative stimuli thus has the potential to alleviate the distress associated with their work by reducing stress and its physiological correlates (i.e., high blood pressure and heart rate in the moment), and longitudinally, by improving the moderator's experience with moderation via association. Awe has also been repeatedly linked to stress reduction specifically (e.g., \cite{Le2019WhenStressor}), but is also linked to more negative affect than cuteness \cite{Le2019WhenStressor,Rankin2020Awe-fullPeriods}. This increased negative affect could potentially heighten moderators’ perception of the negativity in the posts they need to moderate, and this could result in multiple differing outcomes (e.g., counteracting the initial positive affect brought about by the awe stimulus, increasing speed to remove the post faster and reduce the negative affect, etc.). Both awe and cuteness however, have the potential to contribute to an upward spiral in positive affect through repeated exposure (e.g., \cite{Borgi2014BabyChildren,Collado2020ExposureAffect, Hildebrandt1978AdultsCuteness}). Given that these upward spirals could have the power to counterract the negative spirals created by content moderation activities (see \cite{Garland2010UpwardPsychopathology}), both awe and cuteness are worth testing as low-cost, simple interventions that target the moderator as opposed to focusing on the content that needs moderating.

To test these interventions for their suitability for use among content moderators, we ran an experiment on Amazon’s Mechanical Turk (MTurk), a crowdsourced task completion website in which people around the world sign up to participate in small task completions, both academic and commercial, in exchange for payment. This is also a popular platform where a lot of freelance content moderators perform moderation tasks. Participants who are screened for prior content moderation experience were randomly assigned to one of three conditions: taking a break (control), taking a break with pictures of baby animals (cuteness), and taking a break with pictures of beautiful landscapes (awe). Through a simulated content moderation task, in which participants will play the role of a content moderator for a fictional social media platform, we plan to evaluate how effective providing additional positive emotional stimuli is when it comes to reducing a) stress, b) fatigue, and c) emotional distress during breaks in a 45-minute simulated content moderation workday.

\begin{table*}[htbp]
\caption{Summary of Extant Literature's Potential Positive and Negative Effects of Using Awe and Cuteness Induction as Content Moderation Interventions}
\label{table:sum}
    \begin{tabular} {lll}	
    \toprule
      & Positives & Negatives\\ 
    \midrule
    Awe & Stress reduction &  \pbox{6cm}{Potential simultaneous increase in negative affect(self-minimizing perspective).} \\
        &Emotional distress reduction & Increased fatigue over time \\
        &Increased positive affect& \\
        &Increased prosocial tendencies& \\
    \midrule
    Cuteness &  Stress reduction &  Reduction in task completion speed\\
             &  Increased positive affect & \\
             &  Increased desire to protect &\\
             &  Increased desire to nurture &\\
             & Increased task attention &\\
             & Increased care taken in tasks &\\
              
    \bottomrule
    \end{tabular}

\end{table*}

\section{Theoretical Background}
Content moderation literature is clear about the high degree of stress, fatigue, and emotional distress content moderators experience through their work, with many researchers reporting post-traumatic stress disorder (PTSD) among their content moderator participants \cite{Barrett2020ResearchOutsourcing, Dang2018ButModeration,Gillespie2018CustodiansMedia, Riedl2020TheModerators,Wohn2019VolunteerExperience}. In fact, the job is so stressful that it shares common symptomatology with law enforcement workers tasked with finding and capturing child pornographers \cite{Krause2009IdentifyingInvestigators}. This is likely due to the fact that both are regularly faced with similar images: namely, images of suffering victims, descriptions of said suffering, and threats of more suffering to come (see \cite{Krause2009IdentifyingInvestigators, Parks2019DirtyCleaners}). Given the fact that new content is posted on social media at an astronomical rate (see \cite{Internet-live-stats2020InternetStatistics}) Internet Live Stats, 2020 for examples), content moderators’ symptoms are apt to be present for the full duration of a 40-hour work-week, and potentially constantly present, in the event that major disorders like PTSD set in.

To help combat these effects in a low-cost, simple way, we have decided to test the power of awe and cuteness---two positive emotional stimuli---to see if their power to reduce the negative side-effects associated with repeated exposure to negative images \cite{Krause2009IdentifyingInvestigators, Parks2019DirtyCleaners}. Recent research suggests that positive emotional stimuli may have the potential to reduce the negative effects of the job if moderators are exposed at regular intervals \cite{Joye2014AnProsociality,Koh2019ThePossessions,Le2019WhenStressor, Valdesolo2014AweDetection}. In fact, we can already see a current example of cuteness being used to improve the experience of people working for long hours at computers in Google's Colab \footnote{https://colab.research.google.com/}. Programmers can have animated corgis and kittens strut across the top of their screens and meow or bark while they work to add some light entertainment to their user interface. If effective at reducing stress and increasing positive emotions in users, this presents a simple, automated intervention that can be used to supplement existing traditional stress relief techniques, such as mindfulness-based stress reduction (MBSR) meditations (e.g., \cite{Smith2014Mindfulness-BasedStress, Song2015EffectsStudents}). In addition, while MBSR seems to take a minimum of 10 minutes of continuous practice to achieve effects \cite{Innes2018EffectsTrial,Stanley2011Mindfulness-basedCohort}, no such time restrictions have been, to our knowledge, reported for the positive benefits of awe or cuteness, meaning they could theoretically be deployed throughout the workday with minimal disruption to existing schedules. However, despite this potential, to date, no one has actively tested the effects of awe or cuteness for their effectiveness at reducing negative effects and increasing positive affect---the goal of the present work.

\subsection{Awe}
 Awe, and particularly awe experienced as a result of exposure to natural wonders or landscapes, has been repeatedly linked to stress relief, be it pathological (e.g., \cite{Anderson2018AweStudents.}), or due to simple, everyday stressors (e.g., \cite{Joye2014AnProsociality, Koh2019ThePossessions}). This exposure to awesome nature can take the form of actual, in vivo exposure (e.g., \cite{Anderson2018AweStudents.}), or exposure via images or video (e.g., \cite{Koh2019ThePossessions}), as these have both been shown to be effective.

The purported mechanism behind awe's stress reduction properties lies in awe's effect on our self-perception and perception of our stressors \cite{Joye2014AnProsociality, Le2019WhenStressor}. When a person witnesses something that inspires awe, the tendency is for the person to feel small in comparison to whatever it is they are seeing \cite{Bai2017AweSelf}. According to Le and colleagues \cite{Le2019WhenStressor}, this can lead to one of two outcomes, depending on the person's perspective: the awe makes the person's obstacles or challenges feel small and insignificant, or it makes them feel small and insignificant themselves, and unable to meet the challenges ahead. Other research, though, suggests that the perspective adopted may be influenced by the type of awe experienced: positive or negative \cite{Gordon2017TheAwe, Lichtenberg2015Awe:Inhibition}. Awe elicited by a rainbow or a waterfall, for example, does not pose any direct threat to the wellbeing of the average person. Awe elicited from a tornado though, does pose a threat, one that we can feel helpless to face \cite{Lichtenberg2015Awe:Inhibition}. Though all of these stimuli inspire awe, they effect affect and well-being differently \cite{Gordon2017TheAwe}. Thus, extant literature would suggest that exposing stressed people to specifically non-threatening, but awe-inspiring natural stimuli should have stress reduction effects on the person. We thus posit the following hypothesis:

\begin{itemize}
    \item \textbf{H1)} Exposure to awe-inducing stimuli will reduce stress in content moderators compared to the stress levels of content moderators who are not awe-induced.
\end{itemize}

However, experiencing awe may come with a cost in terms of fatigue. This is because the experience of positive awe has been consistently linked to prosociality in study participants \cite{Bai2017AweSelf, Danvers2017GoingDetail,Guan2019AweTendency, Li2019TheAwe,Piff2015AweBehavior,Prade2016AwesHelping,Schneider2017ThePerils}. With fundamentally prosocial work comes what various researchers have dubbed ``compassion fatigue'' \cite{NorrmanHarling2020BreakingFatigue, Neff2020CaringCommunities, Sullivan2019ReducingNurses}. Also called “empathic distress”, compassion fatigue is a cumulative fatigue that care workers experience as a consequence of their employment, as they empathize over time with those under their care \cite{Neff2020CaringCommunities}. Traditionally it is associated with the medical profession (e.g., \cite{NorrmanHarling2020BreakingFatigue,Klein2018QualityFatigue, Neff2020CaringCommunities,Nolte2017CompassionMetasynthesis,Sullivan2019ReducingNurses}), but any job that involves caring for others could theoretically lead to it (e.g., \cite{Lynch2018TheOutcomes,Strohmeier2018FactorsSudan}). This is important to note, as the role of a content moderator is a fundamentally prosocial one: the purpose of the job is to protect the average netizen from the stimuli they themselves are forced to witness. This also means that the more content moderation experience one has, potentially, the more cumulative compassion fatigue they are experiencing, creating worsening symptoms over time. Research in awe has established that seeing an awe-inspiring landscape increases the motivation to perform prosocial behavior; the following question, however, remains unanswered: 

\begin{itemize}
    \item \textbf{RQ1)} Which has a more powerful effect on fatigue: an awe-inspiring stimulus, or accumulated content moderation experience?
\end{itemize}

When it comes to emotional distress, however, awe has a much clearer effect \cite{Koh2019ThePossessions,Le2019WhenStressor,Rankin2020Awe-fullPeriods}. In Koh and colleagues’ \cite{Koh2019ThePossessions} study comparing happiness and awe’s effects on emotional distress, operationalized as negative affect, and their findings indicate that awe has the same distress-reduction properties as happiness. In other words, experiencing awe makes a person feel less negative affect. When distress was operationalized as anxiety, the same effect has been demonstrated \cite{Rankin2020Awe-fullPeriods}. Le and colleagues \cite{Le2019WhenStressor} suggest that the mechanism behind this effect may be how awe makes a person focus on either themselves or others. They call this a “self-distanced” perspective or a “self-immersed” perspective. Their results would indicate that awe inspires a self-distanced perspective in their participants, which consequently made the obstacles in their lives seem less daunting, and the participants less distressed. That said, it is important to note that the studies that talked about distress and awe almost always used distress as an umbrella term to cover a lot of negative emotional experiences that could be treated as distinct in other work. However, there is no evidence to suggest that there is a negative relationship between emotional distress and awe, therefore:

\begin{itemize}
    \item \textbf{H2)} Exposure to awe-inducing stimuli will reduce emotional distress in content moderators compared to the emotional distress levels of content moderators who are not awe-induced.
\end{itemize}

\subsection{Cuteness}

Awe is not, however, the only emotional stimuli that has the potential to reduce the negative effects of content moderation work: the relatively under-studied effect of cuteness has also been shown to have potential when it comes to making difficult situations more bearable \cite{IlhanAslan2016HoldDesigns,Reynolds2011StudyingLibrary}. The sets of literature dedicated to awe and cuteness respectively mirror their topic, with awe literature being vast and well-established, and cuteness literature being young and small by comparison. A major portion of the literature on cuteness actually focuses on cuteness' effect on compassion and care, particularly in relation to mothers and their infants (e.g., \cite{Glocker2009BabyAdults, Kringelbach2016OnBeyond}). This literature stresses that the cuter something is, the more that people – usually mothers (see \cite{Lobmaier2010FemaleFaces,Sherman2013IndividualCareful}) – will be careful around it, and pay attention to it \cite{Nittono2012TheFocus}. Essentially, if something is cute, we want to nurture it and protect it from harm \cite{Glocker2009BabyAdults,Karreman2020ExposureMothers, Sherman2009ViewingCarefulness}. This effect is not relegated to infant humans, though; it has also been shown in animals \cite{Borgi2014BabyChildren, Buckley2016Aww:Cuteness, Little2012ManipulationFaces}, and even objects \cite{Nittono2016TheCuteness,Takamatsu2020MeasuringConstruct}. This desire to protect could have interesting effects for the world of content moderators.

In the case of stress, the admittedly sparse literature that exists would suggest that cute stimuli could have stress reduction properties \cite{Myrick2015EmotionEffect, Takamatsu2020MeasuringConstruct}. Much of this work focuses on animals (e.g., \cite{Myrick2015EmotionEffect,Reynolds2011StudyingLibrary}), and only occasionally separates out the “cute” factor from all the other parts of an animal. Reynolds and Rabschutz \cite{Reynolds2011StudyingLibrary}, for instance, studied exposure to live, in-person animals and how that can reduce stress in university students. Although this intervention was successful, cuteness in particular was not separated out as a specific variable of interest; as the animals in the study were puppies, however, chances are good that they were reasonably cute. Myrick \cite{Myrick2015EmotionEffect}, on the other hand, focused on exposure to animals online – cats, specifically – and found that her participants had a reduced perception of stress. Not only this, but the same participants specified that cuteness was a major part of their motivation to use internet cats as a stress-reduction technique. Although literature specific to cuteness rarely references stress directly (for an exception, see \cite{Takamatsu2020MeasuringConstruct}), work on animals would suggest that exposure to them–-in vivo or otherwise–-should lead to stress reduction. Therefore:

\begin{itemize}
    \item \textbf{H3)} Exposure to cute stimuli will reduce stress in content moderators compared to the stress levels of content moderators who do not see cute stimuli.
\end{itemize}

 In terms of fatigue, cuteness shares the same effect as awe – an increase in the motivation to engage in prosocial behavior (e.g., \cite{Kringelbach2016OnBeyond, Wang2017GettingBehavior})–-in addition to a general desire to care (e.g., \cite{Nittono2012TheFocus}), and so we would expect the same effect of compassion fatigue to occur with cute stimuli. However, when it comes to emotional distress, the evidence is more mixed \cite{Almanza-Sepulveda2018ExploringCuteness,Takamatsu2020MeasuringConstruct}. The majority of studies that discuss cuteness and emotional distress together talk about how higher levels of cuteness make parents particularly empathic and compassionate in regards to the distress of their children (e.g., \cite{Sherman2011CutenessEmotion,Sherman2013IndividualCareful}). This has yet to be demonstrated with other cute stimuli like the baby animals used in the present study, though, nor are the animals in any way presenting distress in the images we procured. In terms of cuteness more generally as a variable of interest, there are only two studies that directly relate it to distress: Almanza-Sepúlveda and colleagues’  study \cite{Almanza-Sepulveda2018ExploringCuteness}, and Takamatsu’s \cite{Takamatsu2020MeasuringConstruct}. In the first of these studies, the measure of cuteness has a significant negative relationship with personal distress \cite{Almanza-Sepulveda2018ExploringCuteness}, but in the second study, there was no relationship at all between the two variables \cite{Takamatsu2020MeasuringConstruct}. Thus, the research would suggest that there is either no relationship or a negative one, but remains inconclusive in terms of which is more likely. All of this taken into account, we posit the following research question and hypothesis:

\begin{itemize}
\item \textbf{RQ2)} Which has a more powerful effect on fatigue: a cute stimulus, or accumulated content moderation experience?
\item \textbf{H4)} Exposure to cute stimuli will reduce emotional distress in content moderators compared to the emotional distress levels of content moderators who do not see cute stimuli.
\end{itemize}

\section{Methods}

The present study takes the form of an experiment with three conditions: control (30-second break every 20 posts moderated, no images), cuteness (30-second break every 20 posts moderated, baby animal images), and awe (30-second break every 20 posts moderated, awe-inspiring landscape images). Participants are randomly assigned to each using SurveyGizmo's built-in tool for random assignment. We used a between-subject design with approximate 120 participants for each condition. This study was reviewed and approved by the institutional review board of a mid-sized American university. We recruited participants from Amazon's Mechanical Turk (MTurk). We first designed a pre-screen survey to ensure that participants had some moderation experience. Participants would receive a ``HIT'' (code) to participate in the online experiment if they qualified.  We monitored the data collection process for about three weeks. We terminated the survey at 387 responses (control 138, awe 123, and cuteness 126), as we had exceeded the total sample size suggested by an a-priori power calculation ran in G*Power seeking a medium effect according to Cohen's conventions (0.25). The projected total sample size at 0.95 observed power with an alpha of 0.05 and three groups was 252, and so to be conservative in case our effect size was smaller than medium (there was, to our knowledge, no existing estimation of effect sizes in content moderator interventions of this type in extant literature), we aimed for a minimum of 300 participants.

\subsection{Participants}
In total, 387 participants aged 23 to 77 (M = 37.78, SD = 10.30) completed the study.  Of these participants, 153 (39.5\%) identified as women, 230 (59.4\%) identified as men, one (0.3\%) as non-binary, and the rest (0.8\%) did not disclose their gender identification. Some (199; 51.4\%) were serving as moderators at the time of participation: 61 for Facebook, 39 on Reddit, 13 on Twitch. tv, 10 on Twitter, 9 on YouTube, 3 on Discord, 2 on Instagram, one on Meetup, one on Slack, one on Telegram, one on Tumblr, and 36 for various websites and forums. Among them, 22 participants were currently serving as content moderators on multiple platforms at once. Of these 199, 160 indicated how long they had occupied their position: 32 (20.0\%) for less than a year, 38 (23.8\%) for one to two years, 31 (19.4\%) for two to three years, 19 (11.9\%) for three to four years, 7 (4.4\%) for four to five years, and 33 (20.6\%) for five or more years. The majority of them (41; 25.6\%) moderated from 5 to 9 hours per week, although 24 (15.0\%) only moderated four hours or less per week. The rest either moderated from 10-14 hours per week (39; 24.4\%), 15-19 hours per week (25; 15.6\%), 20-24 hours per week (17; 10.6\%), 25-29 hours per week (2; 1.3\%), 30-39 hours per week (7; 4.4\%), or 40+ hours per week (5; 3.1\%). 

However, almost all participants (380; 98.2\%) had previously served as a content moderator on some kind of platform. Most of them (126) had previously served as moderators for Facebook, but 93 others on Reddit, 30 on Twitch, 14 on Twitter, 13 on YouTube, 8 on Discord, 3 on Instagram, 2 on 4chan, 2 more on Pinterest, 1 on MySpace, 1 on Meetup, and 87 on various websites and forums. 57 of them served on multiple platforms in the past. Of these 380, 333 (87.6\%) indicated how long they had served in their position: 67 (20.1\%) had served less than a year, 106 (31.7\%) for one to two years, 65 (19.5\%) for two to three years, 39 (11.7\%) for three to four years, 14 (4.2\%) for four to five years, and 43 (12.9\%) for five or more years. The majority (98; 29.4\%) of these moderated from 5 to 9 hours a week, but 46 (13.8\%) moderated four or fewer hours a week. The rest moderated from 10-14 hours per week (70; 21.0\%), 15-19 hours per week (36; 10.8\%), 20-24 hours per week (55; 16.5\%), 25-29 hours per week (7; 2.1\%), 30-39 hours per week (12; 3.6\%), or 40+ hours per week (9; 2.7\%).

\subsection{Procedure}

After providing digital consent to participate in the online experiment, participants are taught what kinds of posts need to be removed – i.e., sexism, racism, threats, etc. – and are given the opportunity to practice on five posts and receive feedback on their choices (were they correct or not, and why; this is only given after the practice rounds, not throughout the study). After successfully completing the five practice rounds, participants reviewed 400 statements and indicated which ones should be removed and which are ok to keep based on the training they just received \footnote{Due to an error in the initial survey design, participants could not be truly randomly allocated using an algorithm. To simulate random allocation, participants were instructed to click one of three links (all with the same title, except Option 1, 2, or 3 appeared afterward) to participate in the study.}.

Throughout the moderation exercise, after every 20 posts (five questions), participants will be given the opportunity to take a 30-second break (as suggested by \cite{Boucsein1997DesignMeasures}, as well as \cite{Lim2016TheTask}), and depending on their condition, the screen is occupied by either the words “Take a break!” (control), a picture of a baby animal and the words “Take a break!” (cuteness), or a picture of a beautiful landscape and the words “Take a break!” (awe). Pictures of baby animals and landscapes were selected based on literature (e.g., \cite{Nittono2012TheFocus,Piff2015AweBehavior}) and were pre-tested by a voluntary sample of personal connections to ensure that they were equally cute or awe-inspiring, depending on the image, and elicited approximately the same levels of positive affect (as tested using Karim et al.’s short-form PANAS  \cite{Karim2011InternationalCultures}). The order of presentation of these images was randomized, but all participants in each condition saw all of that condition’s images. 
After the participant made a decision about the final post (ok or unacceptable), participants were asked to provide their age and gender and expand upon their previous content moderation experience (if any), as well as fill out pre-validated measures of stress \cite{Cohen1983AStress.}, fatigue \cite{Stein2004FurtherForm}, and emotional distress \cite{Kessler1994LifetimeSurvey}.

Finally, participants were thoroughly debriefed and given the option to follow a brief mindfulness-based stress reduction (MBSR) guided meditation \cite{2015MINDFULNESSYouTube}. This treatment has been repeatedly shown to reduce stress in a variety of people and situations (for a full discussion of MBSR and stress, see \cite{Janssen2018EffectsReview}). Participants were also provided with a list of free mental health resources available online, in the event that the meditation is not enough to relieve their emotional or psychological discomfort. At this point, they were given a code that they could enter into the MTurk link they originally clicked to start the study in order to receive their \$12.50 USD compensation for participating in about 45 minutes to one hour.

\subsection{Materials}
\begin{center}
    \jie {Warning: This section may contain offensive content.}
\end{center}
\subsubsection{Stimuli – Edited Tweets}
As previously mentioned, participants simulated moderating 400 edited social media posts. These were initially gathered by undergraduate researchers from actual Twitter posts they found by browsing the social media platform. Undergraduate researchers first read Twitter's community guidelines and a list of banned words as well. Next, they used Google to search for tweets that violated those guidelines or contained these banned words. These tweets were then edited by these same students to remove any identifiable information or additional context. They also developed fake tweets that would be similar to these banned tweets they found online. These posts then were discussed and modified in a lab meeting to ensure that it was commonly understood that these tweets should have been moderated. More specifically, lab members voted on whether or not to ban each tweet, placing themselves in the role of content moderators, much like the present experiment, and only tweets that the majority of lab members considered offensive were marked as banned; others were considered acceptable. Each question is prefaced with a description statement  with six options (5 negative posts and 1 named ``None of these posts need removing''). Below is an example of a typical question in the survey.

``Please select the posts that should be removed according to the training you received. Remember that you can always click away, or choose not to
complete tasks that make you too uncomfortable.''
\begin{itemize}
    \item ``Can't believe the shooter couldn’t kill more people \#iCanDoBetter''
    \item ``Fuck niggers''
    \item ``Gay night club? Just disgusting and I think that gunman did a good job''
    \item ``All Muslims are terrorists''
    \item ``Are all these ching chongs next to me related?''
    \item ``None of these posts need removing''
\end{itemize}
	
\subsubsection{Stimuli – Cute and Awe-Inspiring Images}

At the outset, a group of 40 landscape images and 40 baby animal photos (these two groups were chosen based on extant literature suggesting that these would induce the two desired emotions). These were then piloted on a group of 20 volunteers from the undergraduate population and their families. Participants were asked to rate how cute/awe-inspiring the image was, and also complete the PANAS scale to determine the affect (positive or negative) each image inspired that we might select the images that were the most similar for each condition. The t-test between the target attributes -- cuteness or awe -- revealed no significant difference (t(37) = -0.55, p = .58), meaning that the final selection of images (40 out of 80) are equally cute/awe-inspiring. However, the two t-tests for positivity (t(35) = 9.23, p < .001) and negativity (t(38) = -4.87, p < .001) were both significant, meaning that the cute photos inspired significantly more negative affect, while the awe-inspiring photos inspired significantly more positive affect. The full survey with 400 posts, stimuli about awe and cuteness, and measures, is in the supplementary file.

\subsubsection{Stress}
To measure the stress our participants experienced after performing approximately 45 minutes of simulated content moderation, they completed Cohen and colleagues' perceived stress scale (PSS) \cite{Cohen1983AStress.}. This scale is designed to measure the perceived stress level of a participant at a given point in time, or for a given interval of time. It consists of 10 items that describe common symptoms of stress, and asks participants how often they have experienced said symptom -- for the present study, the time period given was "while you were completing the moderation task" (e.g., "How often have you felt that you were on top of things?"). These were rated on a 5-point Likert scale ranging from 1 (Never) to 5 (Very often). The scale performed reliably, $\alpha$ = .85.

\subsubsection{Fatigue}

To measure participants' levels of fatigue after the experiment, they completed the short form of the Multidimensional Fatigue Symptom Inventory (MFSI-SF), developed by the Moffitt Cancer Center in collaboration with the University of Florida in 1998 \cite{Stein2004FurtherForm}. Similar to the PSS, this measure is designed to capture a person's perceived fatigue at a specific point, or for a given interval, in time. It, too, presents a series of symptoms, and in the present study, participants are asked to indicate how true the statement (including a symptom) is for them at the moment (e.g., "I feel run down", "I am worn out"); this is done using a 5-point Likert scale ranging from 1 (Not at all) to 5 (Extremely). The scale consists of five subscales: general fatigue, physical fatigue, emotional fatigue, mental fatigue, and vigor, which is the opposite of fatigue. Each subscale performed reliably (vigor = .91, general fatigue = .95, physical fatigue = .93, emotional fatigue = .94, mental fatigue = .93) as did the complete scale when treated as a whole, $\alpha$ = .96.

\subsubsection{Emotional Distress}
Finally, to measure emotional distress, participants completed the National Comorbidity Survey's Emotional Distress Scale \cite{Kessler1994LifetimeSurvey}. Again, much like the PSS and MFSI-SF, this measure presents participants with symptoms of emotional distress and asks them to indicate how true each statement is for them at the moment of filling out the measure (e.g., "Worried too much about things", "Frightened"). This is done using a 5-point Likert scale ranging from 1 (Not at all) to 5 (Extremely). This scale also performed reliably, $\alpha$ = .96.

\section{Results}

All analyses were performed in SPSS 26 \cite{IBMCorp.2020IBM27.0}.
To examine the effects of the experimental conditions (control, awe-inspiring stimuli, cute stimuli) and content moderation experience on the fatigue, stress, and emotional distress participants experienced after completing the experiment, we first divided the moderation experience (M = 2.87, SD = 1.60) into two groups by the median, with the “less experienced” group having moderated for two years or less, and the “more experienced” group having moderated for over two years. We then conducted a series of two-way ANOVAs with Tukey post hoc tests to examine the relationships between these variables. 

\subsection{Awe and Cuteness’ Effects on Stress}
For stress, the ANOVA revealed that only moderation experience (F(1,328) = 11.20, p = .001, $\eta^2$ = .03)) had significant effects, with more experienced moderators (M = 2.54, SD = 0.67) feeling more stressed than less experienced moderators (M = 2.30, SD = 0.78), irrespective of the experimental condition (F(2,328) = 2.97, p = .05, $\eta^2$ = .02, observed power = .58), as shown in \autoref{fig:stress}.  We expected that exposure to awe-inducing stimuli would reduce stress in content moderators compared to the stress levels of content moderators who are not awe-induced (H1), which was not supported.  We also anticipated that exposure to cute stimuli would reduce stress in content moderators compared to stress levels of content moderators who do not see cute stimuli (H3), which was also not supported.

\subsection{Awe and Cuteness’ Effects on Fatigue}
\begin{figure}

\begin{subfigure}{0.32\textwidth}
\includegraphics[width=\linewidth]{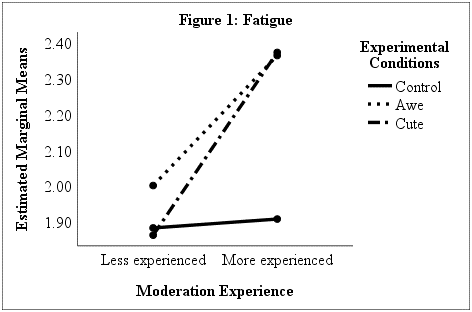}
\caption{Fatigue}
\label{fig:fatigue}
\end{subfigure}
\begin{subfigure}{0.32\textwidth}
\includegraphics[width=\linewidth]{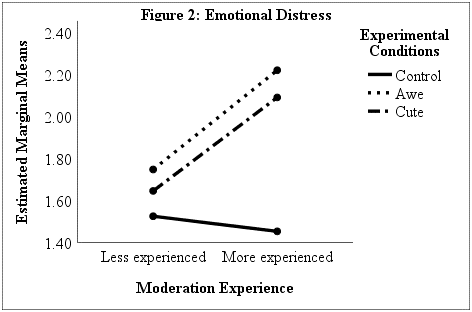} 
\caption{Emotional Distress}
\label{fig:distress}
\end{subfigure}
\begin{subfigure}{0.32\textwidth}
\includegraphics[width=\linewidth]{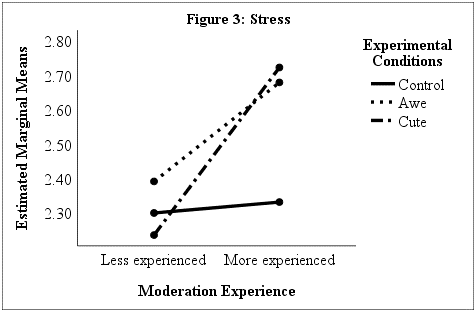} 
\caption{Stress}
\label{fig:stress}
\end{subfigure}
\caption{Interaction effects}
\label{fig:interaction}
\end{figure}

For fatigue, the ANOVA analysis showed that both moderation experience (F(1,328) = 14.00, p <.001, $\eta^2$ = .04)  and experimental conditions (F(2,328) = 5.23, p = .01, $\eta^2$ = .03) had significant effects, but that there was also a significant interaction between these two variables (F(2,328) = 3.52, p = .03, $\eta^2$ = .02).  The significant interaction between moderation experience and condition is visualized in \autoref{fig:fatigue}. The partial eta squared indicated that moderation experience had a stronger impact than experimental conditions on fatigue (RQ1 and RQ2). Simple effects for the interaction revealed that the difference between the three conditions is significant at the more experienced level (F(2,328) = 8.17, p < .001, $\eta^2$ = .05), but not significant at the less experienced level (F(2,328) = 0.57, p = .57, $\eta^2$ = .00, observed power = .14). Specifically, more experienced moderators had a higher fatigue level in the cute (p = .00, d = .47) and awe (p = .00, d = .46) condition than in the control condition, and there was no significant difference in this regard between the cute and awe condition (p = .95, d = .01).

\subsection{Awe and Cuteness’ Effects on Emotional Distress}
For emotional distress, the ANOVA indicated that both moderation experience (F(1,328) = 7.98, p = .01, $\eta^2$ = .02)  and experimental conditions (F(2,328) = 9.92, p <.001, $\eta^2$ = .05) had significant effects, but that there was also a significant interaction effect between these two variables (F(2,328) = 3.51, p = .03, $\eta^2$ = .02). The significant interaction between moderation experience and condition is visualized in \autoref{fig:distress}. More experienced moderators (M = 1.85, SD = 0.99) felt more emotionally distressed than less experienced moderators (M = 1.63, SD = 0.86). Simple effects for the interaction revealed that the difference between the three conditions is significant at the more experienced level (F(2,328) = 12.72, p < .001, $\eta^2$ = .07)  but not significant at the less experienced level (F(2,328) = 0.87, p = .42, $\eta^2$ = .01, observed power = .20). Specifically, more experienced moderators had higher emotional distress in cute (p = .00, d = .64) and awe (p = .00, d = .77) conditions than in the control condition, and there was no significant difference in this regard between cute and awe condition (p = .50, d = .13). Thus, our anticipation that exposure to awe or cute stimuli would reduce emotional distress in content moderators compared to the emotional distress levels of content moderators who do not see cute or awe stimuli (H2 and H4) were not supported.

\section{Discussion}
The goal of the present study was to determine if by adding cute or awe-inspiring stimuli to break times, we could reduce the ill-effects of performing content moderation duties on social media or other online fora. More specifically, we wanted to see if the addition of these images could decrease stress (H1 \& H3), fatigue (RQ1 \& RQ2), and emotional distress (H2 \& H4) after completing an experimental content moderation task. The evidence would suggest that the addition of cute or awe-inspiring stimuli has the opposite effect of our prediction: it actually increases stress levels among experienced content moderators (H1 \& H3); the same is true for emotional distress (H2 \& H4). There are a few possible explanations for this reversed effect. Irrespsective of the exact explanation, this reversed effect could be an indication of content moderation as a unique context, as it seems to contradict findings in other contexts in which positive stimuli can overrule the effects of negative stimuli (e.g., {\cite{Collins2019DigitalRecovery}}). Though we set out to find out how positive emotional stimulation could alleviate the negative side-effects of being a content moderator, what we found instead is that the negativity is so entrenched in some moderators that positive stimulation actually amplifies it, a finding that could have several explanations.

The first possible explanation, also the simplest, is that the addition of any new stimuli during a break disrupts the relaxation that simply ‘being’ can bring (e.g., \cite{Pfeifer2019EnhancedSetting,Yandell1987WastingNothing}). In other words, it could be that, when it comes to stress relief, doing nothing is better than doing something. However, this does not explain why emotional distress only increases significantly among experienced moderators when images are present during breaks, as opposed to creating the same effect across all content moderators. For this effect, we posit the following explanation: the combination of compassion fatigue and increased prosocial tendencies inspired by cuteness and awe. 



As discussed earlier, compassion fatigue is essentially a cumulative effect on wellbeing that occurs in people who work or volunteer in prosocial contexts \cite{Lynch2018TheOutcomes}, such as healthcare (e.g., \cite{Klein2018QualityFatigue} or, in the present study, content moderation. A person experiencing compassion fatigue has spent so much time helping others that they develop a kind of increased susceptibility to taking on others’ problems and struggles, to their own detriment (see \cite{Nolte2017CompassionMetasynthesis}). Although it is called “fatigue”, compassion fatigue is also heavily associated with stress and burnout \cite{NorrmanHarling2020BreakingFatigue,Klein2018QualityFatigue,Neff2020CaringCommunities,Lynch2018TheOutcomes,Monaghan2020CompassionFactors, Sullivan2019ReducingNurses}, which could help explain why it may have affected all of our outcome variables in the present study. Though we anticipated it may have an effect in terms of fatigue (RQ1 \& RQ2), which we appear to have found, as experienced moderators suffered higher levels of fatigue than inexperienced moderators, particularly when exposed to extra stimuli, it would seem that compassion fatigue is all-encompassing in terms of its negative effects. Both cute and awe-inspiring stimuli have been shown to produce elevated prosocial orientation and desire in people \cite{Guan2019AweTendency,Kringelbach2016OnBeyond,Sherman2013IndividualCareful, Wang2017GettingBehavior}, in addition to the positive emotions we anticipated would reduce the negative outcomes we measured \cite{Joye2014AnProsociality,Koh2019ThePossessions,Myrick2015EmotionEffect,Le2019WhenStressor,Takamatsu2020MeasuringConstruct,Valdesolo2014AweDetection}. It would seem that the prosocial effects of these stimuli activate the compassion fatigue of experienced moderators, overriding any positive effects they may have had and instead increasing the stress, fatigue, and emotional distress caused by the desire to help. Although not explicitly measured, theoretically, the content moderator would see a kitten or puppy and be reminded of things like vulnerability and fear and the innocence they are trying to protect in netizens, emphasizing their need to experience the negativity themselves in a self-sacrificial capacity. The longer the content moderator has been at this task, the more emotitonally exhausted they would theoretically become being reminded - by the positive aspects of the world they are trying to protect - of their perceived neverending task and the importance that they continue working at it. Thus, based on our results, it would seem that it is better to do nothing on breaks in professions or volunteer positions where people are at risk of suffering from compassion fatigue.

This is, to our knowledge, the first article to make the association between compassion fatigue and content moderation specifically; if this is indeed the reason positive emotional stimuli ended up producing negative emotional consequences, this means that although cuteness and awe-inspiring images are not the right course of action, there are other options outlined in compassion fatigue research that could also be effective at reducing negative wellbeing effects in the content moderation field. The most popular in extant literature, which comes with a plethora of literature in its own right and was mentioned earlier, is mindfulness-based stress reduction, or MBSR \cite{Janssen2018EffectsReview,Klein2018QualityFatigue}. In fact, we used an MBSR guided meditation as a precaution in this study to ensure that participants did not leave the project depressed or overly anxious. In terms of the specific variables addressed in the present study, Janssen and colleagues \cite{Janssen2018EffectsReview} found MBSR to be an effective way to reduce stress and emotional exhaustion, a concept similar to emotional distress and related to burnout. In addtion, mindfulness is based on the core principle of just being in the moment –- in other words, doing nothing but focusing in silence on what the experience of existence \cite{Querstret2020Mindfulness-BasedMeta-Analysis}. If compassion fatigue is indeed at the heart of our results, then programs like those used by Kneff and colleagues \cite{Neff2020CaringCommunities} and Klein and colleagues \cite{Klein2018QualityFatigue} could be tailored to the content moderation context. If, however, the explanation is that nothing is better than something when it comes to breaks, then MBSR would still be theoretically a helpful practice for content moderators, as it would emphasize the nothingness that brings the relaxation they need. In both cases, however, MBSR would have to be implemented either in longer breaks than what was presented in the present study (see \cite{Innes2018EffectsTrial,Stanley2011Mindfulness-basedCohort}) or as a separate program adjacent to work days. Micro-breaks without additional stimuli could still be beneficial alongside a more structured MBSR program, but this would require additional research to confirm with certainty.

Another possible explanation for the emotional effects we saw in the present study is the idea of affective contrast, which is when the presence of opposite-valence stimuli with affective dimensions amplify each other's emotional effects \cite{Bacon1914AContrast,Harris1929AnContrast}. The baby animals and the beautiful landscapes are designed to evoke positive emotions, but the revised tweets used in the experiment are designed to evoke negative emotions. Some articles discussing affective contrast would suggest that presenting positive stimuli in the midst of negative ones should have actually stabilized or reversed the effects of the negative stimuli (e.g., \cite{Cheng2006NegativeAppraisal}). However, given certain individual differences in terms of optimism and pessimism, it is also possible that people could be more sensitive to contrasts in the negative sense, creating a negative overall affect (e.g., \cite{Geers2002EffectsOptimism-Pessimism}). Similarly, Zhang and Covey \cite{Zhang2014PastConsequences} found that whether an affective contrast came out as a net positive or net negative depended entirely on the person’s preconceptions, which although we can guess were negative given participants' experience with content moderation, we did not explicitly measure in advance. Referring back to the theoretical experience of content moderators exposed to these positive stimuli, this would mean that content moderators who are more experienced are probably framing their task as fruitless and neverending, while the newer content moderators are more motivated to work hard to protect the world's beauty and innocence. In short, although it could be a potential explanation for our findings, the literature is unclear regarding exactly which direction the contrast should go. Future studies could examine this in greater detail by running similar studies including personality and temperament measures, as well as preconceptions regarding content moderation work.

However, another possibility is that we do not see an immediate impact from a cuteness or awe-based intervention because, like compassion fatigue, the effects of positivity are also cumulative. In their study comparing active recovery via a digital game to active recovery via mindfulness practice, Collins et al. \cite{Collins2019DigitalRecovery} found that although essentially equally effective in the short term, the fun stimulation of a game was more effective than the mindfulness program in the long run when it came to increasing energy and decreasing fatigue. This tells us two things: 1) dynamic positive stimulation may be more effective than static, and 2) we may need more exposure to positive stimuli over time to see their ultimate effect. Based on the knowledge we have and our current results, it would seem as though in the short term for micro-break interventions, switching off completely is ideal, particularly for moderators who have already been in the position for many years. However, over time, it is still possible that regular exposure to positive stimuli, particularly if they are dynamic in some way, might be beneficial. To adequately test this long-term possibility, future studies will need to employ a longitudinal design and will also have to test baseline attitudes toward technology and media use (see \cite{Reinecke2016SlackingProcrastination}), as well as pessimism and optimism \cite{Geers2002EffectsOptimism-Pessimism}, in order to tease apart the effects of compassion fatigue, affective contrasts, and simple individual differences to see what is ultimately best for content moderators who wish to make it their permanent career.

\subsection{Practical and Design Implications}
In line with recent work by Steiger et al. \cite{Steiger2021TheModerators} to provide necessary programmatic and technological interventions to offer resources and assistance for mental health, we  provide  several  design interventions that have the potential to protect content moderators or at least mitigate the position's ill effects. Due to the results of positive emotional stimulation being a net negative in the present work, we propose interventions that take the opposite approach and encourage alternative break-time strategies.

The first is a technological intervention that involves how breaks are conducted in a content moderation work or volunteer shift. Although the importance of breaks for any kind of work is well-established in literature (e.g., \cite{Blasche2017EffectsFatigue}), the exact content of these breaks is not always made entirely clear. Although some articles do refer to a ``rest break'' \cite[p.221]{Faucett2007RestTasks}, it is not always made explicit what exactly happens during these breaks. Depending on the real-life work context, a break could consist of browsing social media on one's phone, eating food, or staring at a wall, to name a few possibilities; not all articles specify whether this kind of behavior is prohibited during the assigned breaks or not, and what kinds of break behaviors are beneficial or detrimental to the break's efficacy. As previously stated, our findings would seem to indicate that doing nothing on a break is better than doing something, even if that something is positive, like looking at baby animal photos or beautiful landscapes. We suggest incorporating restful breaks via bots in content moderation user interfaces (UIs). In addition to having a bot remind content moderators to take a 7.5-minute break after every 50 minutes, as proposed by \cite{Boucsein1997DesignMeasures}, the bot should encourage content moderators to remove themselves from the computer, or at the very least from the UI, for the duration of the break. This could even be done automatically, with the UI shutting down or presenting a blank screen and restarting after the designated break time. Of course, preventing content moderators from doing anything during their breaks would be difficult if not impossible remotely, but encouragement to do as little as possible during breaks could increase the effectiveness of said breaks for them. 


The second is a programmatic intervention that could be implemented as  a form of a scheduled rotation of content moderators, to reduce the effects of compassion fatigue . We found that the longer a content moderator had been working (in years), the stronger the negative effects on their stress, emotional distress, and fatigue. This is consistent with compassion fatigue literature, where extended prosocial work leads to burnout or similar symptoms (e.g., \cite{Sullivan2019ReducingNurses}. In the case of medical work, it has been suggested that to get at the root of burnout, the job has to be stripped down to its core, removing any responsibilities outside of providing medical care, in order to reduce the negative effects of being a doctor or nurse \cite{Squiers2017PhysicianDisease}. However, the present study shows that even the most basic of content moderation tasks adds to cumulative negative effects of the position, meaning that there is little that can be 'stripped away'. Thus, we propose extended breaks for content moderators, much like the sabbatical system used by university professors. Sabbaticals have been shown to work to reduce the effects of burnout - which mirror the negative, cumulative effects of content moderation work - in the case of crisis workers, who also traditionally suffer from compassion fatigue \cite{Kang2011TheWell-being}. By implementing 'shifts' of content moderators, the accumulation of ill-effects can be disrupted, allowing for less buildup over time. Future research would have to be conducted to determine the exact duration of a content moderation position before they get rotated off to sabbatical (e.g., one week, one month, two years).


Finally, content moderators should receive systematic training  to learn how to effectively combat the negative effects of their work as a programmatic intervention. While health professionals go through extensive training and years of university to prepare for their jobs, content moderators - especially volunteer content moderators - receive little to no training before they begin work \cite{Seering2019ModeratorAlgorithms}. In addition, while several programs have been developed to help reduce the stress and fatigue of other professionals at risk of compassion fatigue (e.g., \cite{James2018EvaluatingStudy,Sullivan2019ReducingNurses}J), the same is not true of content moderators, who receive only inadequate support from the platform \cite{Dosono2019ModerationCommunities,Roberts2016CommercialWork, Dwoskin2019InsideBattle}. We have already discussed the potential benefits of mindfulness training for content moderators, but other targeted interventions relating to sleep, anxiety, or general physical wellness could also be explored. Now that we know that what content moderators go through could be akin to the experience of crisis workers, police officers, or healthcare professionals, we have many new avenues of promising intervention to validate in the content moderation context.




\subsection{Limitations and Future Directions}
Although this study has important implications for how best to care for content moderators, it is not without limitations. Despite every precaution taken, due to the aforementioned lack of true random allocation, participants could have participated more than once. In addition, since it was an online experiment, it was impossible to monitor exactly what participants did during their breaks. It is possible that they watched YouTube videos or played a video game or did something else during the control conditions or experimental conditions, although the short duration of the breaks would have made that difficult. Future studies could be conducted in the laboratory to better control for these kinds of possibilities. Future studies could also vary the type of stimulus used, as there could be a difference between using static, image stimuli as we did in the present work, and using dynamic video-based stimuli. Finally, although the measures we used covered many of the aspects of compassion fatigue and served as a kind of facsimile for a direct measure of the construct, we did not give our participants a compassion fatigue-specific measure. This is partially because there are only two that we were able to find: The Compassion Fatigue-Short Scale \cite{Yldrm2020TheReliability} and the Nurses’ Compassion Fatigue Inventory \cite{Sabery2017DevelopmentInventory}, both of which are highly tailored to the medical professions. Future work could either create a new, more general compassion fatigue measurement tool, or adapt these existing ones to new contexts to explore the link between compassion fatigue and content moderation in greater depth. Finally, although unlikely, it is possible that participants in one group were feeling particularly distraught prior to starting the experiment. Future studies can measure negative emotion as a pre-test to ensure that this is not the case.

Beyond addressing the aforementioned limitations, future studies could also test some of the relationships and concepts we found here. For instance, we used 30-second breaks between every 20 items moderated, but this is due to the literature on worktime breaks (e.g., \cite{Blasche2017EffectsFatigue}), not due to how breaks currently work for content moderators in the field. Different break time durations, and different spacing between these breaks, should be explored to discover what is optimal for the wellbeing of content moderators. In addition, based on our results, it would seem that mindfulness interventions have the potential to greatly impact content moderators’ experience of their work, but this has yet to be tested directly; it should be compared to other, similar interventions (e.g., transcendental meditation, yoga, relaxing music) to see if it is really the most effective option out there for content moderators.

\section{Conclusions}
In conclusion, although neither awe nor cuteness was found to be efficacious for reducing the negative side effects of performing content moderation, we did uncover some important information when it comes to taking care of the people who protect us on the internet. For one, the effects of stress, fatigue, and emotional distress appear to be cumulative, just like they are in other caring professions; the longer someone serves, the worse they seem to suffer the job’s ill-effects. There is also the strong suggestion of a connection between compassion fatigue and content moderation, giving hope that some of the ways we take care of our healthcare professionals and crisis workers may help alleviate the difficulties of content moderators as well. Instead of alleviating the negative effects of the content moderation work, adding positive stimuli actually seems to have the opposite of the intended effect, leading to new design ideas around work scheduling and technological applications. By continuing to research new and established ways to protect the protectors of the internet, we can help ensure happier, healthier cyberspace for everyone.


\begin{acks}
This research was funded by by National Science Foundation (Award No. 1928627). Thanks to the research assistants in SocialXLab at NJIT for data collection.
\end{acks}

\bibliographystyle{ACM-Reference-Format}
\bibliography{references}






\end{document}